\documentclass[journal]{IEEEtran}
\usepackage{graphics} 
\usepackage{epsfig} 
\usepackage{mathptmx} 
\usepackage{amsmath} 
\usepackage{upgreek}
\usepackage{float}

\begin{document}
%
\title{Active Manipulation of Electromagnetically Induced Transparency in a Terahertz Hybrid Metamaterial}

\author{Tingting Liu
\thanks{T. Liu is with the Laboratory of Millimeter Wave and Terahertz Technology, School of Physics and Electronics Information, Hubei University of Education, Wuhan 430205, People's Republic of China (e-mail: ttliu@hust.edu.cn)}

}

%
\maketitle

\begin{abstract}
The metamaterial analogue of electromagnetically induced transparency (EIT) in terahertz (THz) regime holds fascinating prospects for filling the THz gap in various functional devices. In this paper, we propose a novel hybrid metamaterial to actively manipulate the resonance strength of EIT effect. By integrating a monolayer graphene into a THz metal metamaterial, an on-to-off modulation of the EIT transparency window is achieved under different Fermi levels of graphene. According to the classical two-particle model and the distributions of the electric field and surface charge density, the physical mechanism is attributable to the recombination effect of conductive graphene. This work reveals a novel manipulation mechanism of EIT resonance in the hybrid metamaterial and offers a new perspective towards designing THz functional devices.
\end{abstract}


%
\IEEEpeerreviewmaketitle

\section{Introduction}
\IEEEPARstart{T}{he} electromagnetically induced transparency (EIT) effect has received enormous attention due to the great potential in applications including slow light and nonlinear effects. EIT was first discovered in a three-level atomic system where the destructive quantum interference between a pump and a probe laser beam results in a narrow transparency window within a broad absorption profile\cite{harris1997electromagnetically}. However, the applications of EIT are severely hindered due to the cumbersome experimental conditions involved with the low temperature environment and high intensity lasers. Recently, the advent of metamaterial which possesses the ability to manipulate light-matter interaction in artificially designed structures creates the possibility to mimic the EIT effect in classical optical system\cite{zhang2008plasmon,liu2009plasmonic}. Particularly, the metamaterial analogue of EIT at terahertz (THz) frequencies has been widely investigated since it offers an exciting way to fill the THz gap in  biosensing, optical modulation and slow light devices\cite{singh2009coupling, tao2011recent, liu2012electromagnetically, zhang2013polarization, han2014engineering, manjappa2015tailoring, liang2017plasmonic}.

In practice, it is highly desired to achieve the active manipulation of EIT resonance since it provides more dimensions to the designs and functionalities of metamaterial. Very recently, the integration of active materials into the metamaterial unit cell has been reported as the access route for the realization of dynamically controllable EIT resonance\cite{chen2008experimental, gu2012active,cao2013plasmon,jin2013enhanced,xu2016frequency,manjappa2017active}. For example, Gu et al. integrated the photoconductive silicon into the metal metamaterial and allowed for a giant switching of the EIT resonance under the ultrafast optical pump-terahertz probe measurements\cite{gu2012active}. Cao et al. experimentally demonstrated an amplitude modulation of EIT resonance in a hybrid metamaterial by including a thermally active superconductor resonator and a metal resonator in the unit cell\cite{cao2013plasmon}. The emerging graphene is an active material since its conductivity can be continuously tuned by changing the Fermi level via chemical or electrostatic gating\cite{ju2011graphene,he2015tunable}. Graphene metamaterial provides an alternative platform to realize the active manipulation of EIT resonances\cite{shi2013plasmonic, ding2014tuneable,liu2016actively,yao2016dynamically, zhao2016graphene,xia2016dynamically,he2016terahertz}. However, unlike the above-mentioned hybrid metamaterial, most reported graphene metamaterials shifted the resonance frequency rather than the strength of the EIT resonance, which gives rise to additional noises at adjacent frequency spectra in the modulation process. On the other hand, the nanosctructured graphene in the metamaterial unit cell is in the discrete shape, posing great challenges for the nanoscale fabrication and the tunability implementation in practice.

In this work, we demonstrate an active manipulation of EIT resonance through integrating a monolayer graphene into a THz metal metamaterial. According to the simulation results, an on-to-off modulation of the resonance strength of EIT response can be achieved by shifting the Fermi level of graphene. Based on the two-particle model, the influence of increasing Fermi levels of graphene on the transmission amplitudes of the EIT resonance is theoretically investigated, and an excellent agreement between the theoretical fittings and the simulated results is observed. The distributions of the electric field and surface charge density provide a deeper insight into the modulation mechanism that the active manipulation of the proposed EIT metamaterial is attributed to the recombination effect of the conductive graphene. This work reveals a novel manipulation mechanism of EIT resonance and opens a new perspective towards designing compact and active sensors, slow light devices and switches in THz regime.

\section{Design and simulation of EIT structure}

\begin{figure*} [t]
\label{figure1}
\centering
\includegraphics
[scale=0.6]{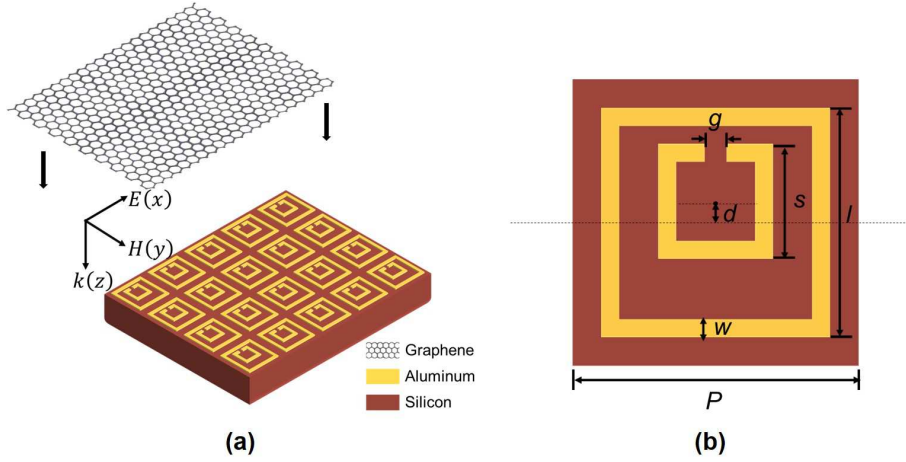}
\caption{\label{fig:epsart} The schematic illustration of the proposed hybrid EIT metamaterial. (a) Three dimensional schematic of the hybrid graphene-metal metamaterial. (b) The top view of the aluminum-based unit cell. The geometrical parameters are as follows: $P= 50$ $\upmu$m, $l= 40$ $\upmu$m, $s= 20$ $\upmu$m, $w=3$ $\upmu$m, $g=4$ $\upmu$m, $d=4$ $\upmu$m.}
\end{figure*}

The schematic of the proposed hybrid EIT metamaterial is illustrated in Fig. 1. The unit cell consists of a split ring resonator (SRR) enclosed within a larger closed ring resonator (CRR) on a semi-infinite silicon substrate. The SRR is square shaped with $s= 20$ $\upmu$m in arm length and $w=3$ $\upmu$m in width, and leaves the gap on the upper side arm with $g=4$ $\upmu$m in length. The CRR also shows square shape and is $l= 40$ $\upmu$m in arm length and $w=3$ $\upmu$m in width. The SRR is positioned close to the upper arm of the CRR with the vertical displacement of $d=4$ $\upmu$m away from the center of the unit cell. Both the SRR and CRR are made of aluminum with a thickness of 200 nm and are periodically arranged with a lattice constant $P=50$ $\upmu$m in both the x and y directions. The resonators are designed to exhibit very close resonance frequencies and highly contrasting resonance linewidths for EIT response in the THz regime. Furthermore, to achieve the active manipulation in the proposed EIT metamaterial, the monolayer graphene is placed on the top of the aluminum-based SRR and CRR.

The numerical simulations with the finite-difference time-domain (FDTD) method are performed using the periodical boundary conditions in the x and y directions and perfectly matched layers absorbing conditions in the z direction along the incident plane wave. In the simulations, the silicon substrate has the refractive index of $n_{Si}=3.42$. The aluminum has the optical constant at THz frequencies described by a Drude model,
\begin{equation}
\label{equation1}
\varepsilon_{Al}=\varepsilon_{\infty}-\frac{\omega_p^2}{\omega^2+i\omega\gamma},
\end{equation}
where the plasmon frequency $\omega_p=2.24\times10^{16}$ rad/s and the damping constant $\gamma=1.22\times10^{14}$ rad/s\cite{ordal1985optical}.The optical conductivity of graphene $\sigma$ modeled within the random-phase approximation consists of interband and intraband contributions. In the lower THz regime, the intraband contribution dominates and the interband contribution is negligible. Without the loss of generality, the graphene conductivity can be derived by a Drude-like model\cite{zhang2015towards,xiao2016tunable,xiao2017strong},
\begin{equation}
\label{equation2}
\sigma_g=\frac{ie^2E_F}{\pi\hbar^2(\omega+i\tau^{-1})},
\end{equation}
where $e$ is the electron charge, $E_F$ is the Fermi level of graphene, $\hbar$ is the reduced Planck’s constant, $\omega$ is the angular frequency. $\tau=\mu E_F/(e\upsilon_F^2)$ is the carrier relaxation time dependent upon the carrier mobility $\mu$ and the Fermi velocity $\upsilon_F$. In our simulations, $\mu=$3000 cm$^2$/V$\cdot$s and $\upsilon_F=1.1\times10^6$ m/s from the experimental measurements are employed\cite{zhang2005experimental, jnawali2013observation}. As can be seen from Eq. (2), shifting the Fermi level $E_F$ enables the dynamic control of the graphene conductivity.

\section{Results and discussions}
To clarify the physical mechanism of the generation of the EIT resonance, three sets of arrays with the aluminum-based unit cell composed of the isolated CRR, the isolated SRR and the proposed EIT structure are investigated. With the plane wave propagation along z direction and the electric field polarized in the x direction, the transmission spectra and field distributions of the three arrays are calculated respectively. The isolated CRR array and the isolated SRR array show independently excited resonances with transmission dips centered around the frequency at about 1.0 THz. As shown in Fig. 2, the isolated CRR array displays a broad resonance with Q factor (defined as the ratio of resonance frequency to the bandwidth at 3 dB) of 1.04 at 1.14 THz. Accordingly, a symmetric distribution of the opposite charges is observed on the left and right arms of the CRR and the electric fields are focused on the two arms perpendicular to the incident electric field. In contrast, the isolated SRR array shows a sharp narrow dip with a Q factor of about 11.88 at 1.08 THz in Fig. 3. With the incident electric field parallel to the gap bearing arm, a circular distribution of the surface charge density along SRR is indicated and the electric fields are concentrated at the two ends of the gap at the resonance frequency. The CRR with broad resonance coupling strongly to the radiation field behaves as the bright mode resonator, while the SRR with narrow resonance coupling weakly to the radiation field serves as the quasi-dark mode resonator. The quite different resonance bandwidths and the close resonance frequency fulfill the design rules of EIT metamaterial.
\begin{figure} [!htbp]
\label{figure2}
\centering
\includegraphics
[scale=0.4]{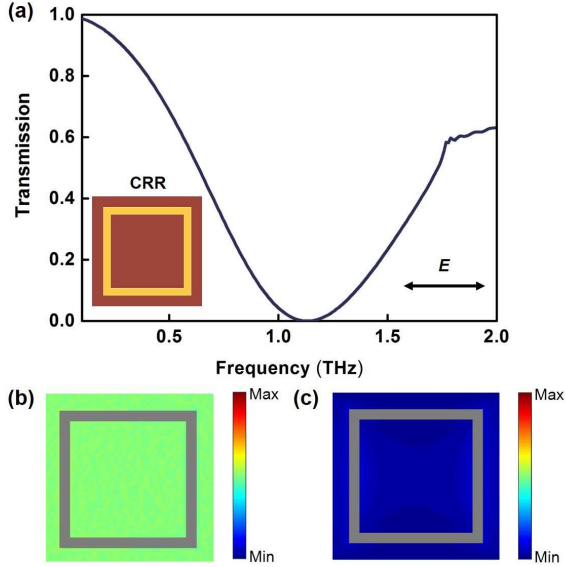}
\caption{\label{fig:epsart} (a) Simulated transmission spectrum, (b) surface charge density distribution and (c) electric field distribution of the isolated CRR array with the electric field polarized in the x direction.}
\end{figure}

\begin{figure} [!htbp]
\label{figure3}
\centering
\includegraphics
[scale=0.4]{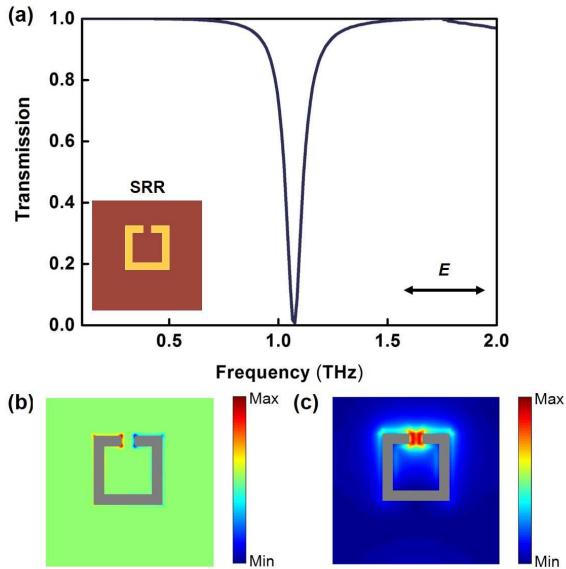}
\caption{\label{fig:epsart}  (a) Simulated transmission spectrum, (b) surface charge density distribution and (c) electric field distribution of the isolated SRR array with the electric field polarized in the x direction.}
\end{figure}

When the aluminum-based CRR and SRR are combined into a unit cell, the EIT resonance is observed with a transparency window at 0.97 THz in Fig. 4. In this case, the CRR and SRR behave as the bright and quasi-dark mode resonators, respectively. The EIT effect due to the near field coupling effect between the bright and quasi-dark modes can be explained by the surface charge density and the electric field distributions of the metamaterial structure. At resonance frequency, the circular distribution of the surface charge density in inner SRR is indicated since the opposite charges accumulate at the two ends of the gap. At the same time, the electric fields concentrate around the SRR gap and the electric fields on the CRR are completely suppressed through the near field coupling effect. As a result, the destructive interference between the two resonators gives rises to the high transparency window of the EIT effect.
\begin{figure} [!htbp]
\label{figure4}
\centering
\includegraphics
[scale=0.4]{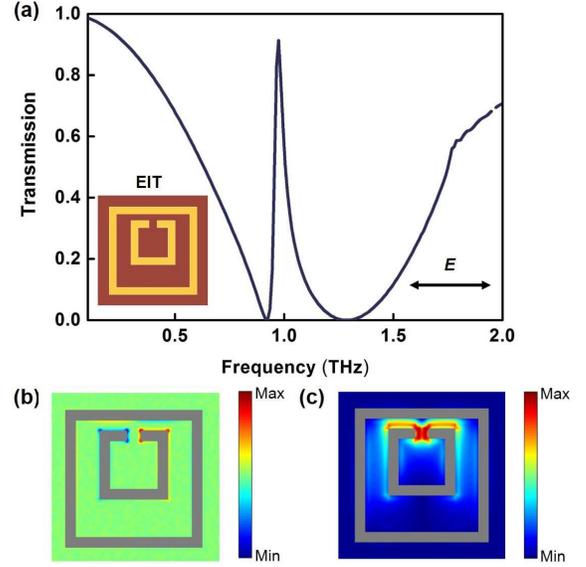}
\caption{\label{fig:epsart}  (a) Simulated transmission spectrum, (b) surface charge density distribution and (c) electric field distribution of the proposed aluminum-based EIT structure with the electric field polarized in the x direction. }
\end{figure}

Next, the active manipulation of the resonance strength of the proposed EIT metamaterial is demonstrated by integrating the monolayer graphene into the metal structure. As shown in Fig. 5, a sharp transparency window with the transmission amplitude of 91.3$\%$ is initially observed at the resonance frequency. When the monolayer graphene is placed on the top of aluminum-based unit cell, the transmission amplitude of the EIT transparency window will experience an on-to-off modulation at the same resonance frequency by shifting the Fermi level of graphene. With the Fermi level as 0.1 eV, the EIT peak has the transmission amplitude of 42.7$\%$. As the Fermi level increases from 0.2 eV to 0.4 eV, the transmission amplitude of the EIT peak sufficiently decreases from 31.3$\%$ to 14.7$\%$. Upon further increase in the Fermi level to 0.8 eV, the EIT transparency window is switched off, leaving a broad resonance dip as low as 4.5$\%$ in the transmission spectrum. Hence, the on-to-off active modulation of the EIT resonance strength is achieved by shifting the Fermi level of integrated monolayer graphene, which highlights better efficiency and feasibility compared with the passive modulation by changing the geometry dimensions in previous investigations.
\begin{figure} [!htbp]
\label{figure5}
\centering
\includegraphics
[scale=0.4]{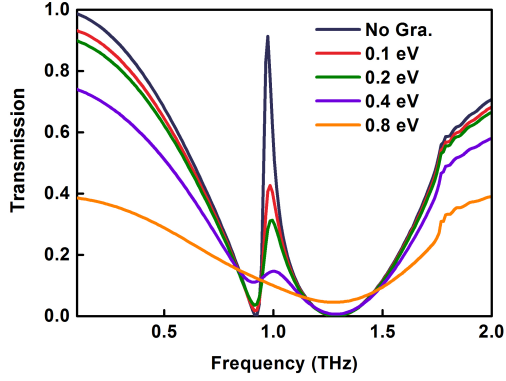}
\caption{\label{fig:epsart} Simulated transmission spectra of the hybrid EIT metamaterial by shifting the Fermi level of the integrated monolayer graphene. }
\end{figure}

The effect of increasing Fermi levels of integrated monolayer graphene on the EIT resonance strength in the proposed metamaterial can be theoretically elucidated with the classical two-particle model. In the EIT structure, the CRR and SRR can be considered as the two particles, i.e. bright and quasi-dark particles, respectively. The near field coupling effect between the two modes under the incident electric fields can be analytically described as\cite{meng2012polarization},
\begin{equation}
\label{equation3}
\begin{split}
    \ddot{x_1}(t)+\gamma_1\dot{x_1}(t)+\omega_1^2x_1(t)+\kappa^2x_2(t)=\frac{q_1E}{m_1},\\
    \ddot{x_2}(t)+\gamma_2\dot{x_2}(t)+\omega_2^2x_2(t)+\kappa^2x_1(t)=\frac{q_2E}{m_2}.
\end{split}
\end{equation}
Here the subscript 1 and 2 represent the bright and quasi-dark particles, respectively. $x_1$ and $x_2$, $\gamma_1$ and $\gamma_2$, $q_1$ and $q_2$, $m_1$ and $m_2$ represent the resonance amplitudes, damping rates, effective charges, effective masses of the bright and quasi-dark modes, respectively. $\omega_1$ and $\omega_2$ are the resonance angular frequencies of the bright and quasi-dark modes. $\kappa$ indicates the coupling coefficient between the bright and the quasi-dark modes. $E=E_0e^{i\omega t}$ represents the incident electric field. To solve the coupled equations, $q_2=q_1/A$ and $m_2=m_1/B$ are substituted in the equations above where A and B are dimensionless constants indicating the relative coupling of the incident radiation with the bright and quasi-dark modes. The resonance amplitudes can be expressed as $x_1=ae^{i\omega t}$ and $x_2=be^{i\omega t}$. Hence, the equation above can be solved as
\begin{equation}
\label{equation4}
\begin{split}
    x_1&=\frac{(\frac{B}{A}\kappa^2+\omega^2-\omega_2^2+i\omega\gamma_2)\frac{q_1E}{m_1}}{\kappa^4-(\omega^2-\omega_1^2+i\omega\gamma_1)(\omega^2-\omega_2^2+i\omega\gamma_2)},\\
    x_2&=\frac{(\kappa^2+\frac{B}{A}(\omega^2-\omega_1^2+i\omega\gamma_1))\frac{q_1E}{m_1}}{\kappa^4-(\omega^2-\omega_1^2+i\omega\gamma_1)(\omega^2-\omega_2^2+i\omega\gamma_2)}.
\end{split}
\end{equation}
Since the linear susceptibility $\chi$ relates the effective polarization of the EIT-like structure $P$ with the incident electric field strength $E$, its expression can be obtained by
\begin{equation}
\label{equation5}
\begin{split}
 \chi=&\frac{P}{\epsilon_0 E}= \frac{q_1x_1+q_2x_2}{\epsilon_0 E}
\\ =&\frac{N}{A^2B}(\frac{A(B+1)\kappa^2+A^2(\omega^2-\omega_2^2)+B(\omega^2-\omega_1^2)}{\kappa^4-(\omega^2-\omega_1^2+i\omega\gamma_1)(\omega^2-\omega_2^2+i\omega\gamma_2)}
\\&+i\omega\frac{A^2\gamma_2+B\gamma_1}{\kappa^4-(\omega^2-\omega_1^2+i\omega\gamma_1)(\omega^2-\omega_2^2+i\omega\gamma_2)},
\end{split}
\end{equation}
where $N$ is the amplitude offset. The real and imagery part of $\chi$ indicate the dispersion and absorption in the structure, and the transmission $T=1-\rm{Im}(\chi)$ can be utilized to fit the simulated transmission spectra in Fig. 5.
\begin{figure} [!htbp]
\label{figure6}
\centering
\includegraphics
[scale=0.4]{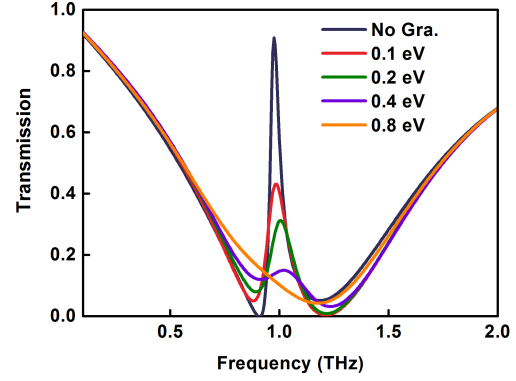}
\caption{\label{fig:epsart}The analytical fitting transmission spectra based on the classical two-particle model for the increasing Fermi level of integrated monolayer graphene. }
\end{figure}

The analytical fitting transmission spectra based on the classical two-particle model are presented in Fig. 6 for different Fermi levels of the monolayer graphene. It can be observed that the fitting spectra exhibit an excellent agreement with the simulated results, demonstrating the validity of the analytical model. In the fitting, the constants $A$ and $B$ with values 40 and 2 imply the interaction between the bright CRR and the incident radiation is 20 times stronger than that of the quasi-dark SRR in the proposed EIT structure. To evaluate the effects of different Fermi levels in the EIT manipulation, the variations of other fitting parameters including $\gamma_1$, $\gamma_2$ and $\kappa$ are depicted in Fig. 7. In the modulation process, the damping rate $\gamma_1$ of the bright CRR, as well as the coupling strength $\kappa$, keeps steady as the Fermi level changes. However, the damping rate $\gamma_2$ of the quasi-dark SRR exhibits a significant increase by nearly two orders of magnitude from 0.005 to 0.48. Thus, active manipulation of EIT resonance is attributed to the increase of the damping rate of the quasi-dark mode.
\begin{figure} [!htbp]
\label{figure7}
\centering
\includegraphics
[scale=0.4]{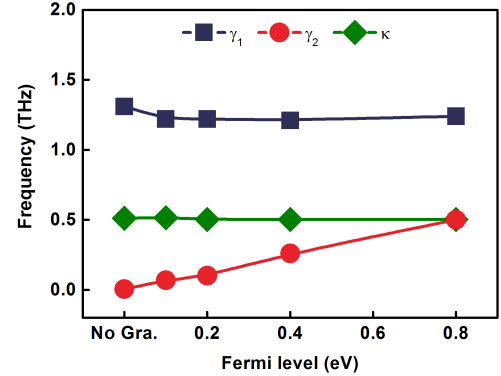}
\caption{\label{fig:epsart} The values of the fitting parameters of the analytical model for different Fermi levels of integrated monolayer graphene.}
\end{figure}

\begin{figure*} [t]
\label{figure8}
\centering
\includegraphics
[scale=0.8]{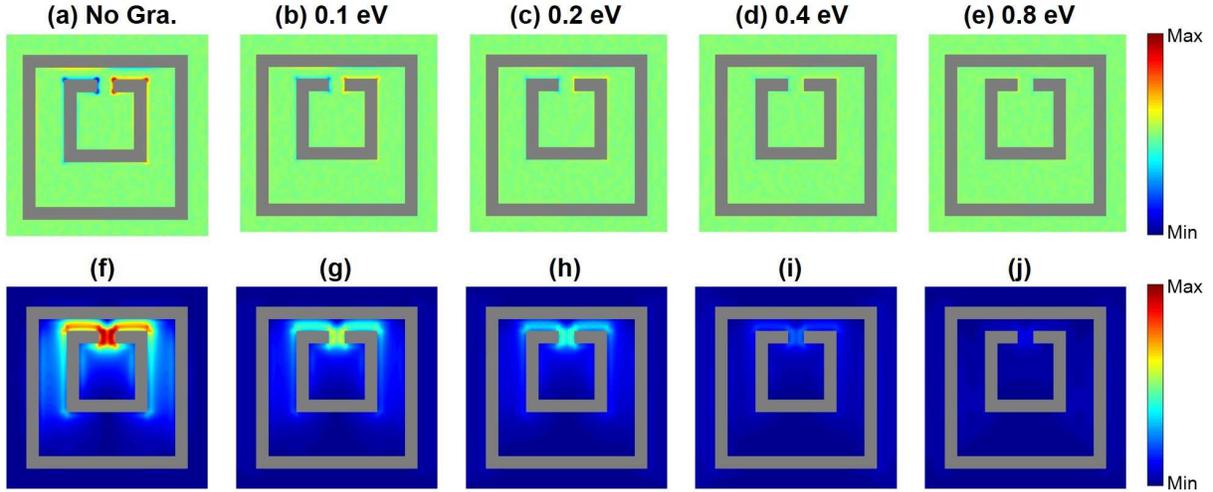}
\caption{\label{fig:epsart} (a) - (e) The simulated surface charge density distributions and (f) - (j) electric field distributions of the proposed hybrid EIT structure at resonance frequency under different Fermi levels of integrated monolayer graphene. }
\end{figure*}

To further explore the physical mechanism of the active manipulation of the EIT resonance, the simulated field distributions of the proposed EIT metamaterial are presented at different Fermi levels in Fig. 8. For the aluminum-based EIT structure in the absence of graphene, the near field coupling effect results in the strong destructive interference between the quasi-dark and bright modes. This effect is reflected in the field distribution diagram at the resonance frequency that the opposite surface charges accumulate and the electric field focuses within the gaps due to the capacitance effect of the SRR gap. At the same time, the damping rate $\gamma_2$ of quasi-dark SRR has very small values, indicating the small resonance loss. In the hybrid EIT structure, the monolayer graphene is integrated with the metal unit cell and behaves as the conductive layer, connecting the gap of the quasi-dark SRR, which leads to the recombination and neutralization of the opposite surface charges. When the Fermi level increases, the conductivity of graphene becomes larger, which further diminishes the capacitance effect at the SRR gap, enhances the loss of the quasi-dark SRR and suppresses the resonance of the quasi-dark mode. The increasing damping rate $\gamma_2$ of the quasi-dark SRR begins to hamper the destructive interference between the bright and quasi-dark modes during the modulation. This process can be observed by the declined surface charges and electric fields around SRR gap. When the Fermi level increases to 0.8 eV, the conductivity of graphene is large enough to completely recombine and neutralize the opposite surface charges of SRR gap. The capacitance effect completely vanishes, leading to the elimination of the strong electric field in SRR gap and leaving the electric field in CRR just as the isolated one. In this case, the damping rate $\gamma_2$ of quasi-dark SRR is very large and the excitation of SRR is completely suppressed, giving rise to the disappearance of EIT peak. Therefore, the underlying physical explanation for the active manipulation of the EIT resonance in the proposed structure lies in the increasing damping rate of the quasi-dark mode resulting from the recombination effect of conductive graphene.

\section{Conclusion}
In conclusion, we have numerically demonstrated the active manipulation of the EIT resonance strength in the graphene-metal hybrid metamaterial. In the metal unit cell composed of quasi-dark SRR enclosed within the bright CRR, the strong near field coupling effect between them gives rise to the EIT resonance in the THz regime. By integrating the monolayer graphene into the metal unit cell, the transmission amplitude of EIT resonance can be dynamically manipulated by shifting the Fermi level of graphene. According to the theoretical analysis based on the classical two-particle model, the active manipulation of EIT effect results from the increasing damping rate of the quasi-dark mode in the proposed structure. Further investigation with surface charge and electric field distributions reveal the physical mechanism of the dynamically controllable EIT resonance lies in the recombination effect of the conductive graphene. This work not only demonstrates the controllable interaction between the monolayer graphene and metal metamaterial, but also holds great potentials in active tunable THz applications.



\end{document}